\documentclass[a4paper]{jpconf}
\pdfoutput=1
\usepackage{graphicx}
\usepackage{url}
\usepackage{dcolumn} %Align table columns on decimal point
\usepackage{bm}      %bold math
\usepackage{lscape}
\usepackage{listings}
\usepackage{verbatim}
\usepackage{bigdelim}
\usepackage{multirow}
\usepackage{bold-extra} % allows bold small caps
\usepackage{amsmath} %it contains the advanced math extensions for LaTeX
\usepackage{amssymb} %it adds new symbols in to be used in math mode
\usepackage[colorlinks=true,linkcolor=blue,citecolor=blue,urlcolor=blue,pagebackref=false]{hyperref}
\usepackage[]{color}
\newcommand{\nonubb}      {$0 \nu \beta \beta$}

\newcommand{\gesevensix}  {$^{76}\textrm{Ge}$}

\newcommand{\mj}          {{\textsc{Majo\-ra\-na}}}
\newcommand{\mjd}         {\textsc{Majo\-ra\-na Demonstrator}}
\newcommand{\dem}         {{\textsc{Demonstrator}}}

\begin{document}
\title{The \mjd: Progress towards showing the feasibility of a tonne-scale \gesevensix\ neutrinoless double-beta decay experiment}

\author{P~Finnerty$^{1,2}$,
E~Aguayo$^{3}$, 
M~Amman$^{4}$, 
F~T~Avignone III$^{5,6}$, 
A~S~Barabash$^{7}$, 
P~J~Barton$^{8}$, 
J~R~Beene$^{6}$, 
F~E~Bertrand$^{6}$, 
M~Boswell$^{9}$, 
V~Brudanin$^{10}$, 
M~Busch$^{11,2}$, 
Y-D~Chan$^{8}$, 
C~D~Christofferson$^{12}$, 
J~I~Collar$^{13}$, 
D~C~Combs$^{14,2}$, 
R~J~Cooper$^{6}$, 
J~A~Detwiler$^{8}$, 
P~J~Doe$^{15}$, 
Yu~Efremenko$^{16}$, 
V~Egorov$^{10}$, 
H~Ejiri$^{17}$, 
S~R~Elliott$^{9}$, 
J~Esterline$^{11,2}$, 
J~E~Fast$^{3}$, 
N~Fields$^{13}$, 
F~M~Fraenkle$^{1,2}$, 
A~Galindo-Uribarri$^{6}$, 
V~M~Gehman$^{9}$, 
G~K~Giovanetti$^{1,2}$, 
M~P~Green$^{1,2}$, 
V~E~Guiseppe$^{18}$, 
K~Gusey$^{10}$, 
A~L~Hallin$^{19}$, 
R~Hazama$^{17}$, 
R~Henning$^{1,2}$, 
E~W~Hoppe$^{3}$, 
M~Horton$^{12}$, 
S~Howard$^{12}$, 
M~A~Howe$^{1,2}$, 
R~A~Johnson$^{15}$, 
K~J~Keeter$^{20}$, 
M~F~Kidd$^{9}$, 
A~Knecht$^{15}$, 
O~Kochetov$^{10}$, 
S~I~Konovalov$^{7}$, 
R~T~Kouzes$^{3}$, 
B~D~LaFerriere$^{3}$, 
J~Leon$^{15}$, 
L~E~Leviner$^{14,2}$, 
J~C~Loach$^{8}$, 
P~N~Luke$^{4}$, 
S~MacMullin$^{1,2}$, 
M~G~Marino$^{15}$, 
R~D~Martin$^{8}$, 
J~H~Merriman$^{3}$, 
M~L~Miller$^{15}$, 
L~Mizouni$^{5,3}$, 
M~Nomachi$^{17}$, 
J~L~Orrell$^{3}$, 
N~R~Overman$^{3}$, 
G~Perumpilly$^{18}$, 
D~G~Phillips II$^{14,2}$,
A~W~P~Poon$^{8}$, 
D~C~Radford$^{6}$, 
K~Rielage$^{9}$, 
R~G~H~Robertson$^{15}$, 
M~C~Ronquest$^{9}$, 
A~G~Schubert$^{15}$, 
T~Shima$^{17}$, 
M~Shirchenko$^{10}$, 
K~J~Snavely$^{1,2}$, 
D~Steele$^{9}$, 
J~Strain$^{1,2}$, 
V~Timkin$^{10}$, 
W~Tornow$^{11,2}$, 
R~L~Varner$^{6}$, 
K~Vetter$^{8,}$\footnote[21]{Alternate Address: Department of Nuclear Engineering, University of California,
Berkeley, CA, USA}, 
K~Vorren$^{1,2}$, 
J~F~Wilkerson$^{1,2,6}$, 
E~Yakushev$^{10}$, 
H~Yaver$^{4}$, 
A~R~Young$^{14,2}$, 
C-H~Yu$^{6}$
and V~Yumatov$^{7}$\\The \mj\ Collaboration}

\address{$^{1}$Department of Physics and Astronomy, University of North Carolina, Chapel Hill, NC, USA}
\address{$^{2}$Triangle Universities Nuclear Laboratory, Durham, NC, USA}
\address{$^{3}$Pacific Northwest National Laboratory, Richland, WA, USA}
\address{$^{4}$Engineering Division, Lawrence Berkeley National Laboratory, Berkeley, CA, USA}
\address{$^{5}$Department of Physics and Astronomy, University of South Carolina, Columbia, SC, USA}
\address{$^{6}$Oak Ridge National Laboratory, Oak Ridge, TN, USA}
\address{$^{7}$Institute for Theoretical and Experimental Physics, Moscow, Russia}
\address{$^{8}$Nuclear Science Division, Lawrence Berkeley National Laboratory, Berkeley, CA, USA}
\address{$^{9}$Los Alamos National Laboratory, Los Alamos, NM, USA}
\address{$^{10}$Joint Institute for Nuclear Research, Dubna, Russia}
\address{$^{11}$Department of Physics, Duke University, Durham, NC, USA}
\address{$^{12}$South Dakota School of Mines and Technology, Rapid City, SD, USA}
\address{$^{13}$Department of Physics, University of Chicago, Chicago, IL, USA}
\address{$^{14}$Department of Physics, North Carolina State University, Raleigh, NC, USA}
\address{$^{15}$Center for Experimental Nuclear Physics and Astrophysics and \\
             Department of Physics, University of Washington, Seattle, WA, USA}
\address{$^{16}$Department of Physics and Astronomy, University of Tennessee, Knoxville, TN, USA}
\address{$^{17}$Research Center for Nuclear Physics and Department of Physics, Osaka University, Ibaraki, Osaka, Japan}
\address{$^{18}$Department of Physics, University of South Dakota, Vermillion, SD, USA}
\address{$^{19}$Centre for Particle Physics, University of Alberta, Edmonton, AB, Canada}
\address{$^{20}$Department of Physics, Black Hills State University, Spearfish, SD, USA}

\ead{paddy@unc.edu}

\begin{abstract}
  The \mjd\ will search for the neutrinoless double-beta decay (\nonubb) of the \gesevensix\ isotope with a mixed array of enriched and natural germanium detectors.  The observation of this rare decay would indicate the neutrino is its own anti-particle, demonstrate that lepton number is not conserved, and provide information on the absolute mass-scale of the neutrino. The \dem\ is being assembled at the 4850 foot level of the Sanford Underground Research Facility in Lead, South Dakota. The array will be contained in a low-background environment and surrounded by passive and active shielding. The goals for the \dem\ are: demonstrating a background rate less than 3~t$^{-1}$~y$^{-1}$ in the 4 keV region of interest (ROI) surrounding the 2039 keV \gesevensix\ endpoint energy; establishing the technology required to build a tonne-scale germanium based double-beta decay experiment; testing the recent claim of observation of \nonubb\ \cite{Kla06}; and performing a direct search for light WIMPs (3-10 GeV/c$^{2}$). % This article will provide an overview of the \mjd\ focusing on the sensitivity of a light WIMP search.
\end{abstract}

\section{Introduction}
\begin{figure}[htbp]
  \centering
      \includegraphics[width=0.75\textwidth]{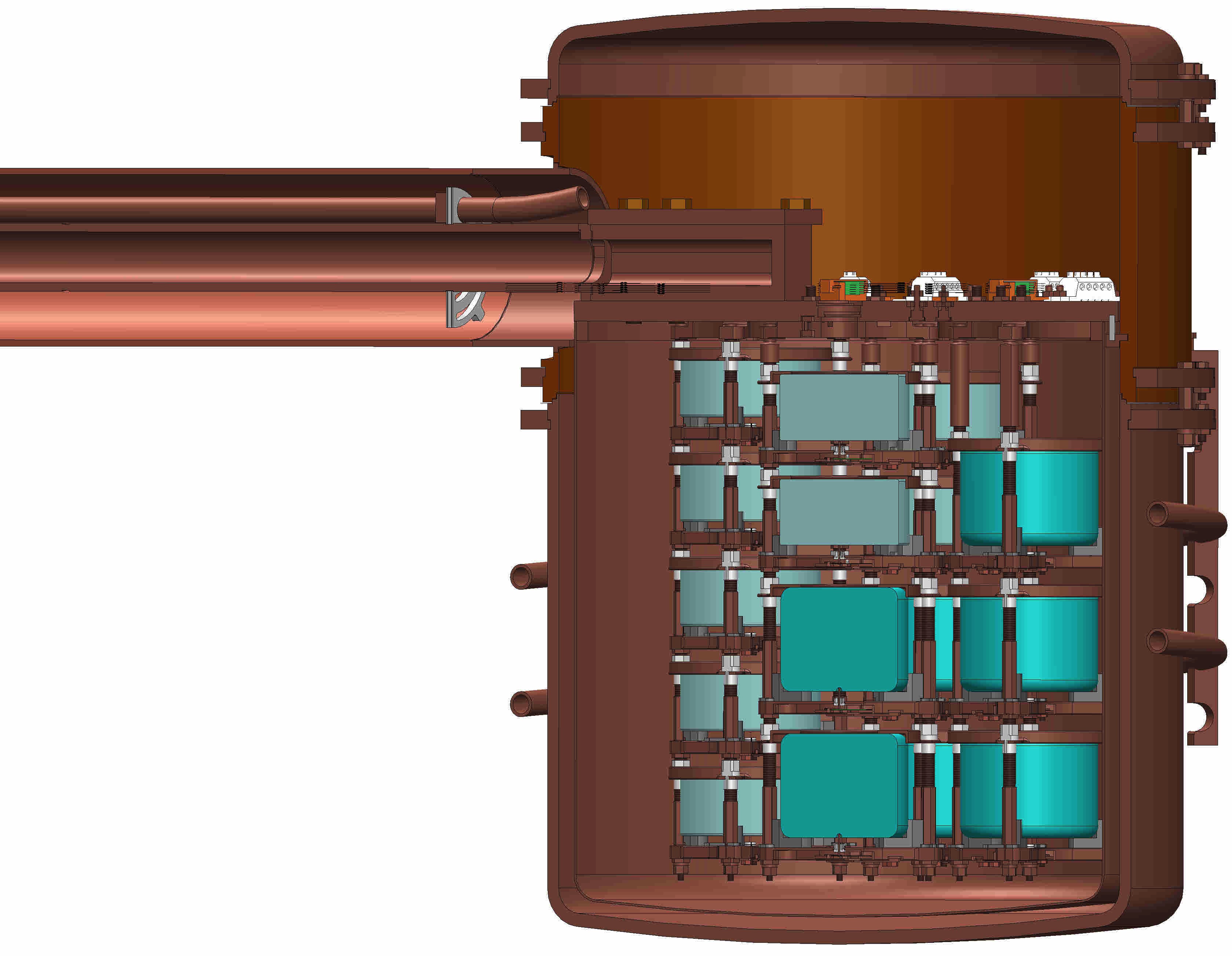}
\caption{A cross sectional view of a \mjd\ cryostat.  The strings within the cryostat hold a mixture of natural (smaller/light blue) and enriched (larger/dark blue) germanium detectors (Color online).\label{fig:cryo_internals}} 
\end{figure}

\begin{figure}[htbp]
  \centering
      \includegraphics[width=0.75\textwidth]{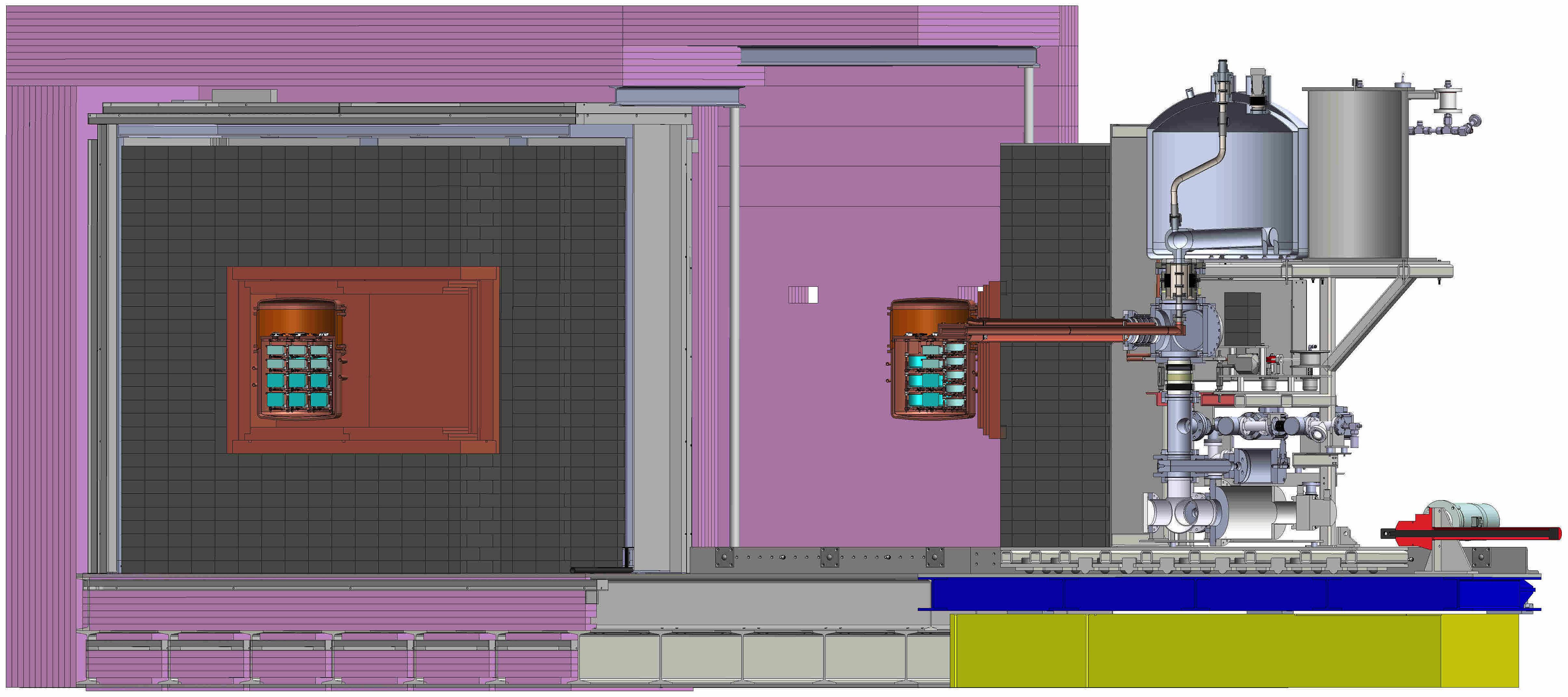}
\caption{The \mjd\ is shown here with both active and passive shielding in place.  One cryostat is in place inside the shield while the other is being positioned for insertion.  For scale, the inner copper shield is 20'' high and 30'' in length (Color online). \label{fig:mjd}} 
\end{figure}

The \mj\ collaboration \cite{Agu11,Wil12,Sch12} will search for the neutrinoless double-beta decay (\nonubb) of \gesevensix. The observation of this rare decay would indicate the neutrino is its own anti-particle, demonstrate that lepton number is not conserved, and provide information on the absolute mass-scale of the neutrino (see Ref.~\cite{Ver12} for a recent review of \nonubb\ theory). Reaching the neutrino mass-scale associated with the inverted mass hierarchy, $20 - 50$ meV, will require a half-life sensitivity on the order of 10$^{27}$~y.  This corresponds to a signal of a few counts or less per tonne-year in the \nonubb\ peak (2039 keV for \gesevensix). To observe such a rare signal, one will need to construct tonne-scale detectors with backgrounds in the region of interest at or below $\sim$1~t$^{-1}$~y$^{-1}$.  The \mj\ collaboration is constructing the \dem, an array of high-purity germanium (HPGe) detectors at the 4850 foot level of the Sanford Underground Research Facility (SURF) in Lead, South Dakota. The \dem\ will consist of a mixture of natural (10-15 kg) and $>$86\% enriched \gesevensix\ (up to 30 kg) HPGe detectors in two low-background cryostats.  Each cryostat will contain seven closely-packed stacks, called strings, and each string will have up to five crystals (see Figs.~\ref{fig:cryo_internals}~and~\ref{fig:mjd}).  The \dem\ aims to: 
\begin{enumerate}
  \item demonstrate a background rate less than 3~t$^{-1}$~y$^{-1}$ in the 4 keV region of interest (ROI) surrounding the 2039 keV \gesevensix\ endpoint energy;
  \item establish the technology required to build a tonne-scale germanium based \nonubb\ experiment;
  \item test the recent claim of observation of \nonubb\ \cite{Kla06};
  \item and perform a direct search for light WIMPs (3-10 GeV/c$^{2}$). 
\end{enumerate}

The \mj\ and GERDA \cite{Sch05} collaborations are working together to prepare for a single tonne-scale \gesevensix\ experiment that will combine the best technical features of both experiments.  There are two main differences between the experimental techniques employed by \mj\ and GERDA.  First, the \dem\ array will be deployed in a custom vacuum cryostat, whereas GERDA is submerging theirs in liquid argon.  Secondly, the \dem\ will use a compact shield with lead, oxygen-free copper, electroformed copper, and scintillator paddles, whereas GERDA is using the liquid argon and high-purity water as a shield.

\section{Detector Technology}
% PPC detectors
The \mj\ collaboration will use p-type point contact (PPC) HPGe detectors.  These detectors \cite{Luk89,Bar07} have been demonstrated to provide both exceptional energy resolution ($<$2.0 keV FWHM at 1332 keV, $<$4.0 keV FWHM at 2039 keV) and low-energy thresholds ($\sim$500 eV) \cite{Bar07,Aal08}, see e.g. Figure~1 in Ref.~\cite{Aal08}.  Several successful prototypes have been commercially produced and successfully operated in an underground environment \cite{Aal08,Bar09,Aal10,Aal11,Mar10}.  The PPC detectors used in the \dem\ will each have a mass of 0.6-1.0~kg.

\section {Background Mitigation Techniques}
% backgrounds - pure materials & eforming
One of the technical goals of the \dem\ is to show that a 1~t$^{-1}$~y$^{-1}$~ROI$^{-1}$ background rate is achievable for a tonne-scale experiment; this scales to 3~t$^{-1}$~y$^{-1}$~ROI$^{-1}$ background rate for the \dem, which is $\sim$100 times lower than previous Ge experiments.  One of the primary methods for achieving this rate is to deploy the detectors inside two independent cryostats that minimize the amount of interstitial material (see Figs.~\ref{fig:cryo_internals}~and~\ref{fig:mjd}).  In addition, the materials used to fabricate the \dem\ have been screened and selected based on strict radiopurity requirements.  The main structural material closest to the detectors within the \dem\ is electroformed copper due to its intrinsically low-background and excellent physical properties.  Electroformed copper is made by electroplating from high-purity commercial oxygen-free copper nuggets onto stainless steel mandrels.  The electroforming process greatly reduces the radioactivity due to U, Th and cosmogenically-produced $^{60}$Co.  Electroformed copper is being used to fabricate both of the cryostats for the \dem. A prototype cryostat has been built from commercial oxygen-free copper and is currently being tested.  The successful performance of the prototype cryostat will give the \mj\ collaboration confidence in the future performance of both electroformed copper cryostats.

% backgorunds - psa
Further background reduction in the \dem\ is achieved using the technological features of the PPC detectors.  A \nonubb\ event will deposit all of its energy within a $\sim$1 mm$^3$ region inside the PPC detector.  These are called single-site events (SSE).  In contrast, gamma-rays from radioactive contaminants of sufficient energy to affect the 2039 keV ROI will typically Compton scatter several times with a scattering length of $\sim$1 cm. These are called multi-site events (MSE).  The \dem\ will reduce Compton scattered gamma-ray backgrounds by implementing sophisticated pulse shape analysis (PSA) techniques to separate SSE from MSE within a single Ge crystal.  In addition to signal PSA, background events that deposit energy in more than one detector can be removed with an anti-coincidence cut.

% backgrounds - cosmogenics
Cosmogenically-produced isotopes represent another source of background.  Cosmic rays can create $^{60}$Co in copper components as well as $^{68}$Ge, $^{60}$Co and tritium within the detectors themselves.  In an effort to minimize these backgrounds, the surface exposure of detectors and copper components will be minimized.  Also, all copper components within the cryostats are being electroformed underground to reduce the amount of $^{60}$Co activation. Reducing surface exposure is the only option for most cosmogenically-produced isotopes.  However, the background due to $^{68}$Ge is a special case.  $^{68}$Ge decays to $^{68}$Ga which has a half-life of 67.71 minutes.  The low-thresholds and excellent energy resolution of PPC detectors allow for a single-site time-correlation (SSTC) cut, which looks forward and backward in time from the current event in the ROI to search for signatures of parent or daughter isotopes \cite{Eji04,Eji05}. The \dem\ will implement a SSTC cut for the $^{68}$Ge-$^{68}$Ga coincident decay for five $^{68}$Ga half-lives following the 10 keV K-shell de-excitations.  The SSTC method can also be used to reduce backgrounds due to $^{208}$Tl (T$_{1/2}$ = 3.05 min) and $^{214}$Bi (T$_{1/2}$ = 19.9 min) in the Ge crystals and the inner mount.  The SSTC and SSE/MSE cuts will both be used in order to maximize background reduction. 

\section{Background Modeling Tools}
The \mj\ collaboration uses a Monte Carlo simulation framework, MaGe \cite{Bos11}, which is co-developed by the \mj\ and GERDA collaborations, to estimate background contributions to the ROI.  MaGe is based on the GEANT4 \cite{Ago03,All06} and ROOT \cite{Bru97} packages and provides a common set of physics processes, event generators, and various management classes.  MaGe also provides a user-interface for output classes and detector geometries.  The radioactivities of all components within the \dem\ have been measured.  The results of these measurements are used with MaGe to develop a comprehensive background model for the \dem.

\section{Dark Matter}
The low-energy performance of PPC detectors, due to their low-capacitance point-contact design, makes them suitable for dark matter (DM) searches.  The direct detection of DM, which makes up $\sim$23\% of the Universe, remains an active area of research. Weakly interacting massive particles (WIMPs), a generic class of potential DM candidates, are widely regarded as the most promising candidate \cite{Ber12}.  Recent direct searches for WIMPs have hinted towards a low-mass WIMP at $\sim$7~GeV/c$^2$ \cite{Aal08,Aal10,Aal11,Mar10,Ang11}.  The favored $\sim$7~GeV/c$^2$ WIMP is also consistent with the theoretical model proposed by Hooper~and~Goodenough~\cite{Hoo11} to explain the excess flux of gamma-rays from the Galactic center.  With a 500 eV threshold, the sensitivity of the \dem\ to a 7~GeV/c$^2$ WIMP is expected to reach $10^{-43}$~cm$^2$ (normalized to nucleon) allowing it to test these recent claims \cite{Bar09,Mar10,Gio12}.

As a part of the research and development (R\&D) efforts for the \dem, the \mj\ collaboration has deployed a custom PPC detector (MALBEK) in the Kimballton Underground Research Facility (KURF) in Ripplemead, Virginia, at a depth of 1450 meters water equivalent \cite{Fin10}.  MALBEK is a 465~g natural Ge crystal housed in a low-background cryostat and shield.  The data acquisition (DAQ) and slow-control system for MALBEK was built around ORCA, an object oriented data acquisition that provides an easy to use, graphical interface for manipulation of experimental hardware and data streams \cite{How04}.  The goals for this R\&D effort are to:
\begin{enumerate}
  \item test a \mjd-like DAQ;
  \item validate the \mjd\ background model;
  \item systematically characterize the low-energy spectrum;
  \item and perform a light WIMP search.
\end{enumerate}

\section{\dem\ Implementation}
The \mjd\ prototype cryostat is expected to be completed by late 2012 and will contain two strings of natural Ge PPC crystals.  The electroformed cryostats, Cryostat 1 and Cryostat 2, are being fabricated in a phased approach.  Cryostat 1 will contain seven strings of Ge crystals, four holding natural Ge crystals and three containing Ge crystals enriched to greater than 86\% \gesevensix.  Cryostat 1 is expected to be completed towards the end of 2013.  Cryostat 2 will contain seven strings of Ge crystals, all of which enriched to greater than 86\% \gesevensix.  Cryostat 2 is expected to be completed by the end of 2014.  The \mjd\ will search for \nonubb\ and low-mass WIMPs once fabrication is completed at SURF.  In addition, the \mjd\ should be able to verify or refute the recent observational claim of \nonubb\ \cite{Kla06} within 2-3 years of commissioning Cryostat 1.

\section{Acknowledgements}
We acknowledge support from the Office of Nuclear Physics in the DOE Office of Science under grant numbers DE-AC02-05CH11231, DE-FG02-97ER41041, DE-FG02-97ER41033, DE-FG02-97ER4104, DE-FG02-97ER41042, DE-SCOO05054, DE-FG02-10ER41715, and DE-FG02-97ER41020. We acknowledge support from the Particle and Nuclear Astrophysics Program of the National Science Foundation through grant numbers PHY-0919270, PHY-1003940, 0855314, and 1003399. We gratefully acknowledge support from the Russian Federal Agency for Atomic Energy. We gratefully acknowledge the support of the U.S. Department of Energy through the LANL/LDRD Program. N. Fields is supported by the DOE/NNSA SSGF program. G. K. Giovanetti is supported by the Department of Energy Office of Science Graduate Fellowship Program (DOE SCGF), made possible in part by the American Recovery and Reinvestment Act of 2009, administered by ORISE-ORAU under contract no. DE-AC05-06OR23100.

\section*{References}

\bibliography{PSF_MJD_PASCOS}

\end{document}